\documentclass[a4paper,11pt]{article}
\usepackage{pos}
\usepackage{mathtools}
\usepackage{mathrsfs}
\usepackage{amsmath}
\usepackage{amsfonts}
\usepackage{soul}

\title{Importance of Multiplicity Fluctuations in Entropy Scaling}
\author*[a]{Patrick Carzon}
\author[a]{Matthew D. Sievert}
\author[a]{Jacquelyn Noronha-Hostler}

\affiliation[a]{Illinois Center for Advanced Studies of the Universe, Department of Physics, University of Illinois at Urbana-Champaign, Urbana, IL 61801, USA}

\emailAdd{pcarzon2@illinois.edu}
\emailAdd{msievert@illinois.edu}
\emailAdd{jnorhos@illinois.edu}

\abstract{One of the greatest uncertainties in heavy-ion collisions is the description of the initial state. Different models predict a wide range of initial energy density distributions based on their underlying assumptions. Final flow harmonics are sensitive to these differences in the initial state due to the nearly linear mapping between eccentricities and anisotropic flow harmonics. The Trento code uses a model-agnostic approach by phenomenologically parameterizing the initial state and constraining those parameters from a Bayesian analysis. There the multiplicity fluctuations were determined by a one parameter $\Gamma$ distribution. However, initial-state models arising from the Color-Glass Condensate (CGC) framework lead to an initial energy density which is outside the functional form considered in Trento and its later Bayesian analyses because they rely on log-normal multiplicity fluctuations. We compare $T_{A}T_{B}$ scaling (CGC-like) to $\sqrt{T_{A}T_{B}}$ scaling (preferred from a Trento Bayesian analysis) and find that the $T_A T_B$ form together with log-normal fluctuations is a reasonable candidate to describe the multiplicity fluctuations but leads to larger eccentricities, which would affect the extraction of viscosity in small systems.
}

\FullConference{%
  HardProbes2020\\
  1-6 June 2020\\
  Austin, Texas}

\begin{document}
\maketitle

\section{Introduction}
Understanding the initial state of heavy-ion collisions requires real-time non-perturbative  calculations in Quantum Chromodynamics that cannot be done on the lattice. The initial condition affects the final state, which affects the extraction of viscosities from comparisons to experimental measurements of flow harmonics $v_n$.  The initial state can be quantified using eccentricities $\varepsilon_{n} = | \langle e^{i n \phi} \rangle |$, which are strongly correlated with $v_n$ \cite{Gardim:2011xv,Gardim:2014tya}. The extraction of viscosity is directly correlated with the model's $\varepsilon_{n}$ \cite{Luzum:2009sb}. Thus, the correct description of the initial state is crucial to extract properties of the Quark Gluon Plasma. Recently a Bayesian analysis used a phenomenologically based initial state, TRENTO, and demonstrated a preference for the initial entropy density $s \propto \sqrt{T_{A}T_{B}}$ where $T_A , T_B$ are the nuclear thickness functions   \cite{Moreland:2018gsh}. The Bayesian analysis assumed a $\Gamma$ functional form for the event-by-event multiplicity fluctuations \cite{Moreland:2018gsh}. This choice restricts the range of models considered, excluding potentially viable models such as a Color-Glass Condensate (CGC) description of the initial state \cite{Nagle:2018ybc, Lappi:2006hq, Chen:2015wia, Romatschke:2017ejr}. In Ref.~\cite{Nagle:2018ybc} the authors investigated a CGC-like linear scaling of the initial energy density $\epsilon \propto T_A T_B$ and considered a log-normal distribution for the functional form of the multiplicity fluctuations. Here, we systematically study the impact of these two initial-state models and choices of multiplicity fluctuations on the initial-state eccentricities.

\section{Methods}
The initial state in Trento is characterized by thickness functions of the form $T_{A, B} = \omega_{A, B} \int dz \, \rho$
% \begin{equation} \label{eq:thickness}
%     T_{A,B}(x,y)=\omega_{A,B}\int dz \, \rho_{A,B}(x,y,z) .
% \end{equation}
where $\rho$ is the number density of individual nucleons and $\omega$ is a multiplicity weight which fluctuates event by event. The weights $\omega$ are sampled from a distribution whose mean in $1$ but allow for high-multiplicity fluctuations within a given nucleon. In Trento, that distribution is chosen to be a one-parameter $\Gamma$ distribution of the form 
\begin{equation} \label{eq:gammaflucs}
    P_{k}(\omega)=\frac{k^{k}}{\Gamma(k)}\omega^{k-1}e^{-k\omega},
\end{equation}
 with the parameter $k$ controlling the shape of the distribution \cite{Moreland:2018gsh}. In the limit $k \rightarrow \infty$, the distribution Eq. (\ref{eq:gammaflucs}) approaches a delta function $\delta(\omega)$; it becomes wider as $k\rightarrow1$; and blows up at 0 when $k<1$. Given this assumed functional form, the Bayesian analysis found that the initial-state model $s \propto \sqrt{T_{A}T_{B}}$ was preferred. Other models \cite{Nagle:2018ybc}, using instead a  log-normal distribution
\begin{equation} \label{eq:lognormflucs}
    P_{k}(\omega)=\frac{2}{\omega k\sqrt{2\pi}}e^{-\frac{\ln^{2}(\omega^{2})}{2k^{2}}},
\end{equation}
in conjunction with the linear scaling $\epsilon \propto T_A T_B$ can also describe the data. The log-normal distribution Eq. (\ref{eq:lognormflucs}) becomes a delta function $\delta(\omega)$ in the limit $k \rightarrow 0$, while large values $k \sim \mathcal{O}(1)$ correspond to a wider distribution.

\begin{figure}[ht]
    \centering
    \includegraphics[width=0.65\textwidth]{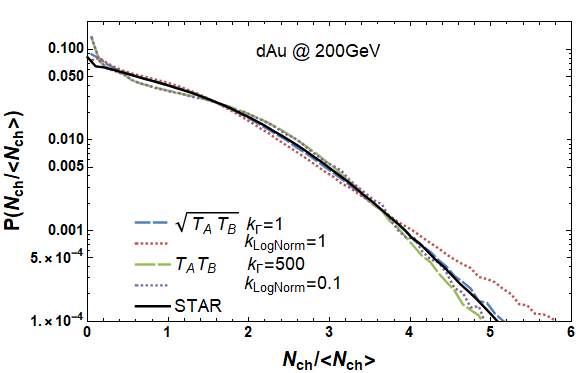}
    \caption{Multiplicity Distribution of dAu for functional forms $\sqrt{T_{A}T_{B}}$ and $T_{A}T_{B}$ using best fits for $\Gamma$ and lognormal multiplicity fluctuation distributions.}
    \label{fig:dAuMultiplicity}
\end{figure}
To systematically study the effects of these different choices, we have added both linear scaling and the log-normal distribution to Trento. The effect of these choices can be seen in the multiplicity distributions shown in Fig. \ref{fig:dAuMultiplicity}, where $k$, for each distribution and functional form, is tuned to best reproduce STAR data for dAu at 200 GeV. Below $N_{ch}/\langle N_{ch}\rangle = 4$, there is good agreement with the data in all four cases. The high-multiplicity tail of the distribution is where the curves start to depart from the data, with $\sqrt{T_{A}T_{B}}$ and log-normal fluctuations overpredicting the data by a significant amount. $\Gamma$ multiplicity fluctuations work well with Trento's $\sqrt{T_{A}T_{B}}$ and log normal fluctuations do not. Across both fluctuation distributions, $T_{A}T_{B}$ favors a narrow distribution providing few fluctuations while $\sqrt{T_{A}T_{B}}$ prefers a wider distribution leading to a lot of fluctuations, suggesting linear scaling is able to match data by mean field while $\sqrt{T_{A}T_{B}}$ requires many fluctuations to reach the same result.

\section{Results}
\begin{figure}[ht]
    \begin{center}
    \includegraphics[width=0.48\textwidth]{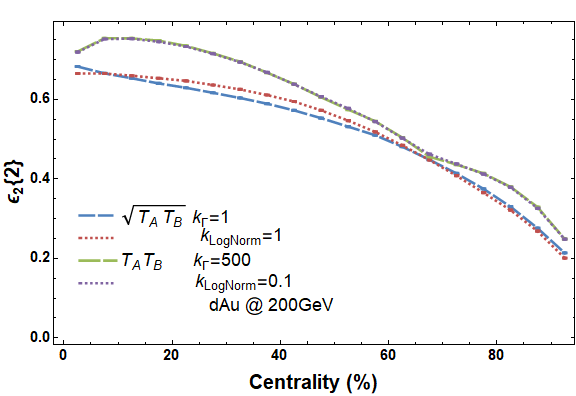}
    \includegraphics[width=0.48\textwidth]{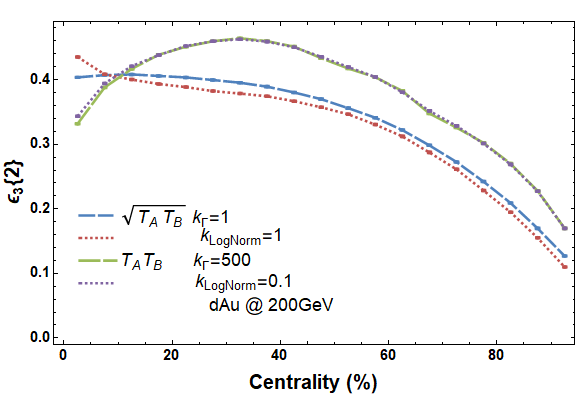}
    \caption{Two particle eccentricities of dAu for functional forms $\sqrt{T_{A}T_{B}}$ and $T_{A}T_{B}$ using best fits for $\Gamma$ and lognormal multiplicity fluctuation distributions.}
    \label{fig:eccs}
    \end{center}
\end{figure}

The effect of these choices in functional form and multiplicity fluctuations can be seen in the plots of $\varepsilon_{2}\{2\}$ (left), that describes the ellipticity of the initial state, and $\varepsilon_{3}\{2\}$ (right), which characterizes the triangular geometry, shown in Fig. (\ref{fig:eccs}).
Despite agreement in multiplicity distribution below $N_{ch}/\langle N_{ch}\rangle = 4$, the eccentricities differ significantly. Linear scaling leads to larger eccentricities than $\sqrt{T_{A}T_{B}}$, except for a dip in $\varepsilon_{3}\{2\}$ in central events. This could be a mean field effect, $\sqrt{T_{A}T_{B}}$ spreads out the energy more than $T_{A}T_{B}$ and slightly washes out the geometry. The difference in choice of multiplicity fluctuations is negligible except for $\sqrt{T_{A}T_{B}}$ at low centralities, where the $\Gamma$ fluctuations trend downward or level off while the log normal trend upward. The effect of the differences in the eccentricities of the two functional forms will be seen in the extraction of shear viscosity with $T_{A}T_{B}$ needing a larger viscosity.

\section{Conclusion}
Trento's preferred functional form $T_R = \sqrt{T_A T_B}$ works best with a $\Gamma$ multiplicity fluctuation distribution, suggesting the choice of functional form is correlated with the multiplicity fluctuation distribution. The linear functional form $T_R \propto T_A T_B$, which is outside the scope considered in the Trento Bayesian analysis, is also able to reproduce the experimental dAu multiplicity distributions with fewer multiplicity fluctuations. This may suggest that the parameter space considered in Trento's Bayesian was overly restrictive, excluding some viable models in small systems. Between the two models, there is a noticeable difference in the magnitude of the eccentricities $\varepsilon_2 \{2\} , \varepsilon_3 \{2\}$, although the trends are qualitatively similar. The difference in eccentricities suggests that there is systematic uncertainty in the Trento Bayesian extraction of QGP viscosities in small systems, which could be controlled by increasing the functional space of the Bayesian analysis.

\section*{Acknowledgements}

The authors acknowledge support from the US-DOE Nuclear Science Grant No. DE-SC0019175, the Alfred P. Sloan Foundation, and the Illinois Campus Cluster, a computing resource that is operated by the Illinois Campus Cluster Program (ICCP) in conjunction with the National Center for Supercomputing Applications (NCSA), and which is supported by funds from the University of Illinois at Urbana-Champaign.
 
%\cite{Luzum_2009}

\bibliographystyle{JHEP}
\bibliography{bibliography}

\end{document}